\documentclass[,nofootinbib,amsmath,amssym,preprint,aps]{revtex4}
\usepackage{hyperref,slashed,color}
\usepackage{graphicx,subfig}
\begin{document}
 \draft
\title{$Z'$ Boson Mixings with $Z\!-\!\gamma$ and Charge Assignments}

\author{Ying Zhang$^1$,   Qing Wang$^{2,3}$\footnote{Corresponding author at:
Department of Physics, Tsinghua University, Beijing 100084,
P.R.China\\ {\it Email
address}:~wangq@mail.tsinghua.edu.cn (Q.Wang).}}

\address{$^1$School of Science, Xi'an Jiaotong University, Xi'an, 710049, P.R.China\\
    $^2$Center for High Energy Physics, Tsinghua University, Beijing 100084, P.R.china\\
    $^3$Department of Physics,Tsinghua University,Beijing 100084,P.R.China}

\date{May 17, 2009}

\begin{abstract}
Based on the general description for $Z'\!-\!Z\!-\!\gamma$
mixing as derived from the electroweak chiral Lagrangian, we characterize and classify the various
new physics models involving the $Z'$ boson that have appeared
in the literature into five classes: 1. Models with minimal $Z'\!-\!Z$
mass mixing; 2. Models with minimal $Z'\!-\!Z$ kinetic mixing;
3. Models with general $Z'\!-\!Z$  mixing; 4. Models with
$Z'\!-\!\gamma$ kinetic and $Z'\!-\!Z$ mixing; and 5. Models with
Stueckelberg-type mixing. The corresponding mixing matrices are
explicitly evaluated for each of these classes. We constrain and classify the $Z'$ boson
charges with respect to quark-leptons
by anomaly cancellation conditions.

\end{abstract}
\maketitle


\vspace{1cm}
\section{Introduction}

With the running of the LHC
at CERN Geneva, a TeV energy era
begins and researchers are anxiously expecting a possible new revolution in
particle physics. There are various predictions from both the Standard Model (SM)
and new physics
models. Among these the appearance
of possible new underlying interactions beyond conventional
strong/weak/electromagnetic gauge interactions is of special
interest. From knowledge accumulated in resent years in particle physics,
we know that the expected new interactions at least must govern the electroweak symmetry breaking
that result in the
massive $W^{\pm}$ and $Z^0$ bosons and  may further be responsible
for the origin of masses for ordinary quarks and leptons. Theoreticians have
also touted various ambitious alternative sources of
these new interactions, such as unifications,
supersymmetries, and extra dimensions. With the exception of the well-known
scalar-type interactions which suffer unnaturalness, triviality
and hierarchy problems, the typical new interaction that avoids the
shortcomings of elementary scalar fields
is a gauge interaction and
minimal such kind of interaction involves an additional so-called $U(1)'$ gauge
interaction. In most instances this extra $U(1)'$ gauge force is a
"relic" of some larger underlying new physics gauge interactions
such as those occurring in GUT models, string theories, left-right
symmetric models and models deconstructed from extra space
dimensions. Alternatively, in some special models, the $U(1)'$ gauge force
takes on a special role: for example 1) in little Higgs type models, it can partially
remove the quadratic divergence from the SM Higgs mass at the one
loop level\cite{RizzoARXIV2006}; 2) in topcolor-assisted technicolor
(TC2) models, it ensures top quark condensation while not for the bottom
quark \cite{Hill,Lane,Chiv}; 3) in SUSY models, it can
mediate SUSY breaking\cite{Z'SUSY}; and 4) in models based on string
theory, it mediates particles  communicating between the hidden and
visible sectors \cite{CasselARXIV2009}. This represents but a sampling of
new physics models involving additional $U(1)'$ factors: a recent
review of others can be found in Ref.\cite{Langacker}.

Phenomenologically, we are interested in the possibility of experimentally finding
the carrier, an
electrically-neutral color singlet spin-one boson $Z'$,
of this additional gauge force especially at the LHC.
As a detection has not been made so far, this boson has to be massive and
the corresponding $U(1)'$ gauge symmetry must be violated. The more
preferred and exciting experimental finding would be that the $Z'$ mass is
relatively light compared with the other new physics particles, for then
it might arise as a first signature of the new physics
beyond SM at the LHC. This prospect heightens the need for theoretical
studies of such a light $Z'$ boson and its interactions with known particles
would also be of the special importance in new physics research.

Physically, one main effect of the $Z'$ boson
derives from its mixings with conventional $Z$ boson
and $\gamma$ photon; another stems from its gauge couplings to ordinary quarks and leptons, which
leads to various charge assignments. There exist a diversity of
new physics models involving the $Z'$ boson,
each model has its own arrangement of $Z'-Z-\gamma$ mixings and
$Z'$ coupling to ordinary quarks and leptons.
To compare models, a model independent investigation is needed of
these Z' boson interactions with known particles, particularly in classifying and
comparing the role of the Z' boson within each model.
The electroweak chiral Lagrangian (EWCL) method provides such
a platform to perform model independent research. In our previous
paper \cite{Z'our}, we have written down the bosonic part
up to order $p^4$ of the most genral EWCL involving the $Z'$
boson\footnote{In the Lagrangian, terms involving a  neutral Higgs boson that only plays a passive role are
also included to help in matching unitarity requirements within the theory.} and
known particles.
This EWCL alos describes the most general  $Z'\!-\!Z\!-\!\gamma$ mixings,
and with it we can further classify the various $Z'\!-\!Z\!-\!\gamma$
mixings that appear in each model enabling us to compare and
discriminate between the different new physics models\footnote{It should be
emphasized that a $p^4$ order EWCL provides some special degrees of
freedom for the $Z'\!-\!Z\!-\!\gamma$ mixings. For example, all
kinetic mixings are from $p^4$ order terms in EWCL (see Eq.(\ref{LNK})),
as a $p^2$ order EWCL only cannot offer the most general
$Z'\!-\!Z\!-\!\gamma$ mixings. }. Here the classification categorizes
the general $Z'\!-\!Z\!-\!\gamma$ mixings into several
simplifying cases that appear in the new physics models in the
literature. The reason in doing this is because the general
$Z'\!-\!Z\!-\!\gamma$ mixings is too complex to be discussed
analytically, while we will show that for all simplifying cases presented
in this paper, mixings can be diagonalized exactly. This
improves on the approximate diagonalization result usually used in the
literature and we can exhibit explicitly the relationship between
the various simplifying cases. The main purpose of this paper is
to present these finding s and moreover to generalize
the EWCL given in Ref.\cite{Z'our} to include  the $Z'$ boson
couplings to ordinary quarks and leptons for the most general
charge assignments. In terms of these charges, new physics models
involving the $Z'$ boson can also be classified. Because most of the
experimental searches for the $Z'$ boson depend heavily on these
charge assignments and on how  $Z'$ mixes with $Z$ and $\gamma$,
we combine a discussions on these two issues in present paper.

This paper is organized as follows. In
Sec.\ref{SEClagrangian}, we first give a short review of the
bosonic part of the EWCL involving the $Z'$ boson and general
$Z'\!-\!Z\!-\!\gamma$ mixings. In Sec.\ref{MixingModels}, we
classify the various models involving the $Z'$ boson that have appear in the
literatures according to their arrangements of the
$Z'\!-\!Z\!-\!\gamma$ mixings.
 In Sec.\ref{SECcharge}, we set up the general
$Z'$ boson charge assignments to the ordinary quarks and leptons
 in terms of the anomaly cancellation conditions. Sec.\ref{Sum} provides a summary of the paper.

\section{The Bosonic part of the EWCL involving the $Z'$ boson and $Z'\!-\!Z\!-\!\gamma$
mixings}\label{SEClagrangian}

As given in Ref.\cite{Z'our}, the covariant derivative in the EWCL including the $Z'$ boson is
\begin{eqnarray}
D_\mu\hat{U}=\partial_\mu\hat{U}+igW_\mu\hat{U}-i\hat{U}\frac{\tau_3}{2}g'B_\mu
-i\hat{U}(\tilde{g}'B_\mu+g^{\prime\prime}X_\mu)I\;,\label{DUdef}
\end{eqnarray}
where the two by two unitary field $\hat{U}$ represents four
Goldstone boson degrees of freedom resulting from spontaneous
symmetry breaking of $SU(2)_L\otimes U(1)_Y\otimes
U(1)'\rightarrow U(1)_{em}$, and $\tilde{g}'$ is a
Stueckelberg-type coupling constant associated with which is a special kind of $U(1)$.
To help in understanding this choice of covariant
derivative, we denote $SU(2)_L\otimes U(1)_Y\otimes U(1)'$ group
elements as $(e^{i\theta^at^a_L+i\theta't'},e^{i\theta t})$ for which the
Hermitian matrices $t^a_L$ ($\theta^a$) with $a=1,2,3$, $t$ ($\theta$) and
$t'$ ($\theta'$) are generators (group parameters) of $SU(2)_L$,
$U(1)_Y$ and an extra $U(1)'$ respectively. The electromagnetic
$U(1)_\mathrm{em}$ group generator has now been generalized from its
traditional expression to $t_\mathrm{em}\equiv t_L^3+t+ct'$ depending on an
additional arbitrary parameter $c$. This generator results in
the $U(1)_\mathrm{em}$ group element
$(e^{i\theta_\mathrm{em}(t^3_L+ct')},e^{i\theta_\mathrm{em}t})$ and
we can label the representative element for the corresponding coset by
$(\hat{U},1)$. Group theory tells us that each symmetry breaking generator
corresponds to a coset which can be represented by introducing a
representative element for each coset. Denoting the
representative element by $n$, its transformation rule to
$n'$ under the action of an arbitrary group element $g$ is then $gn=n'h$ where $h$
is an element belonging to the un-broken subgroup. Specifically for the above gauge
group, this transformation rule then stipulates that
\begin{eqnarray}
(e^{i\theta^at_L^a+i\theta't'},~e^{i\theta
t})(\hat{U},1)\stackrel{gn=n'h}{=====}(\underbrace{e^{i\theta^at_L^a+i\theta't'}
\hat{U}e^{-i\theta(t_L^3+ct')}}_{\hat{U}'},~1)
\underbrace{(e^{i\theta(t_L^3+ct')},~e^{i\theta
t})}_{U(1)_\mathrm{em}}
\end{eqnarray}
which yields the following transformation rule for the Goldstone field $\hat{U}$ under
$SU(2)_L\otimes U(1)\otimes U(1)'$
\begin{eqnarray}
\hat{U}'=e^{i\theta^at_L^a+i\theta't'}~\hat{U}~e^{-i\theta(
t_L^3+ct')}\;.\label{Utrans1}
\end{eqnarray}
 The choice of the Goldstone field in the two dimensional internal space
corresponds in taking the generator $t_L^a=\tau^a/2$, $t=t'=1$
(Note, according to our arrangement of group elements,
$t$ and $t'$ act on different spaces, so $t=t'=1$ will not cause confusion).
With (\ref{Utrans1}) and the standard
$SU(2)_L\otimes U(1)_Y\otimes U(1)'$ transformation rule for
electroweak gauge fields $W_\mu,B_\mu$ and the extra $U(1)'$ gauge
field $X_\mu$, we derive the action of the covariant derivative on the
Goldstone field $\hat{U}$ as: $D_\mu\hat{U}=\partial_{\mu}
\hat{U}+i(gW_{\mu}+g_XX_\mu)\hat{U}
-i\hat{U}(\frac{\tau^3}{2}g'+cg')B_\mu$.
Further identifying
$g_X\equiv -g"$ and $cg'\equiv\tilde{g}'$, we obtain the result
given in Eq.(\ref{DUdef}). With symmetry breaking pattern $SU(2)_L\otimes
U(1)_Y\otimes U(1)'\rightarrow U(1)_{em}$, the Higgs mechanism
ensures that the Goldstone bosons represented by the $\hat{U}$ field
will be eaten out by the electroweak gauge bosons $W^{\pm},Z^0$ and
$Z'$ which then acquire mass. Here $W_\mu$, $B_\mu$ and $X_\mu$ are respectively the
gauge fields of $SU(2)_L$, $U(1)_Y$ and $U(1)'$ before mixing.

The full bosonic part of the Lagrangian up to order $p^4$ is
\begin{eqnarray}
\mathcal{L}_{Stueck-SU(2)_L\otimes U(1)_Y\otimes
U(1)'\rightarrow
U(1)_{em}}=\mathcal{L}_0+\mathcal{L}_2+\mathcal{L}_4\;,
\end{eqnarray}
with each term in the Lagrangian defined as
\begin{eqnarray}
\mathcal{L}_0&=&-V(h)\;,\\
\mathcal{L}_2 &=&\frac{1}{2}(\partial_\mu h)^2
-\frac{1}{4}f^2\mathrm{tr}[\hat{V}_\mu\hat{V}^\mu]
+\frac{1}{4}\beta_1f^2\mathrm{tr}[T\hat{V}_\mu]\mathrm{tr}[T\hat{V}^\mu]
+\frac{1}{4}\beta_2f^2\mathrm{tr}[\hat{V}_\mu]\mathrm{tr}[T\hat{V}^\mu]\nonumber\\
&&+\frac{1}{4}\beta_3f^2\mathrm{tr}[\hat{V}_\mu]\mathrm{tr}[\hat{V}^\mu]
+\beta_4f(\partial^{\mu}h)\mathrm{tr}[\hat{V}_\mu]\;,\label{L2}\\
\mathcal{L}_4&=&\mathcal{L}_K+\mathcal{L}_B+\mathcal{L}_H+\mathcal{L}_A\;,
\end{eqnarray}
where $T\equiv\hat{U}\tau_3\hat{U}^\dag$ and $\hat{V}_\mu\equiv
(\hat{D}_\mu\hat{U})\hat{U}^\dag$. Here the Higgs field $h$ is treated as
$p^0$ order and
\begin{eqnarray}
\mathcal{L}_K&=&
    -\frac{1}{4}B_{\mu\nu}B^{\mu\nu}
    -\frac{1}{2}\mathrm{tr}[W_{\mu\nu}W^{\mu\nu}]
    -\frac{1}{4}X_{\mu\nu}X^{\mu\nu}\;\nonumber\\
\mathcal{L}_B
&=&\frac{1}{2}\alpha_1gg'B_{\mu\nu}\mathrm{tr}[TW^{\mu\nu}]
+\frac{i}{2}\alpha_2g'B_{\mu\nu}\mathrm{tr}[T[\hat{V}^\mu,\hat{V}^\nu]]
+i\alpha_3g\mathrm{tr}[W^{\mu\nu}[\hat{V}^\mu,\hat{V}^\nu]]+\ldots\nonumber\\
\mathcal{L}_H &=&(\partial_\mu
h)\Big\{\alpha_{H,1}\mathrm{tr}[T\hat{V}^\mu]\mathrm{tr}[\hat{V}_\nu\hat{V}^\nu]
    +\alpha_{H,2}\mathrm{tr}[T\hat{V}_\nu]\mathrm{tr}[\hat{V}^\mu\hat{V}^\nu]
    +\alpha_{H,3}\mathrm{tr}[T\hat{V}_\nu]\mathrm{tr}[T[\hat{V}^\mu,\hat{V}^\nu]]+\ldots\Big\}\;.\nonumber
\end{eqnarray}
All coefficients in above Lagrangian are functions of Higgs field $h$.
Detailed expressions can be found in Ref.\cite{Z'our}.

Mixings among $Z'\!-\!Z\!-\!\gamma$ come from the gauge boson mass
term $\mathcal{L}_M$ and kinetic term $\mathcal{L}_K$. In the unitary
gauge $\hat{U}=1$, they  become
\begin{eqnarray}
\mathcal{L}_{M,Z'-Z-\gamma}&=&\frac{f^2}{8}(1\!-\!2\beta_1)(gW^3_\mu\!-g'B_\mu)^2
+\frac{f^2}{2}(1\!-\!2\beta_3)(g^{\prime\prime}X_\mu\!+\tilde{g}'B_\mu)^2
    \nonumber\\
    &&+\frac{f^2}{2}\beta_2(g^{\prime\prime}X_\mu+\tilde{g}'B_\mu)(gW^{3,\mu}-g'B^\mu)\;,\label{LNM}\\
\mathcal{L}_{K,Z'-Z-\gamma}&=&
    -\frac{1}{4}B_{\mu\nu}B_{\mu\nu}
    -\frac{1}{4}X_{\mu\nu}X^{\mu\nu}
     -\frac{1}{4}(1\!-\!\alpha_8g^2)(\partial_\mu W^3_\nu\!-\partial_\nu W^3_\mu)^2\label{LNK}
   \\
   &&+\frac{1}{2}\alpha_1gg'B_{\mu\nu}(\partial_\mu W^3_\nu\!-\partial_\nu
    W^3_\mu)+gg^{\prime\prime}\alpha_{24}X^{\mu\nu}(\partial_\mu W^3_\nu-\partial_\nu W^3_\mu)
    +g'g^{\prime\prime}\alpha_{25}B_{\mu\nu}X^{\mu\nu}\;.\nonumber
\end{eqnarray}
Apart from the four gauge couplings
$g,g',g^{\prime\prime},\tilde{g}'$, seven extra dimensionless
parameters $\beta_1,\beta_2,\beta_3$ and
$\alpha_1,\alpha_8,\alpha_{24},\alpha_{25}$ determine the mixing
terms. Of these eleven, $\alpha_8$ can be
absorbed into the redefinition of field $W^3_\mu$ and coupling constant $g$
by
\begin{eqnarray}
W^3_\mu\rightarrow
\frac{W^3_\mu}{\sqrt{1-\alpha_8g^2}}\hspace{2cm}g\rightarrow
g\sqrt{1-\alpha_8g^2}\;.
\end{eqnarray}
Hence we are left with ten parameters, and on eliminating the three gauge couplings
$g,g',g^{\prime\prime}$, leaves us {\it seven independent} parameters
$\tilde{g}',\beta_1,\beta_2,\beta_3,\alpha_1,\alpha_{24},\alpha_{25}$
that are related to mixings. However, the mixing masses and kinetic terms given by
(\ref{LNM}) and (\ref{LNK}) are so complex that to diagonalize them
we must exploit the general $3\times 3$ rotation matrix $U_{ij}$
\begin{eqnarray}
(W^3_\mu,~B_\mu,~X_\mu)^T= U(Z_\mu,~A_\mu,~Z'_\mu)^T\;,\label{Udef}
\end{eqnarray}
which has nine matrix elements. The fact that no correction terms arise for the
kinetic terms $-\frac{1}{4}B_{\mu\nu}B_{\mu\nu}$ and
$-\frac{1}{4}X_{\mu\nu}X^{\mu\nu}$ leads to two constraints on the matrix
elements of $U$,
\begin{eqnarray}
(U^{-1,T}U^{-1})_{22}=(U^{-1,T}U^{-1})_{33}=1\;,
\end{eqnarray}
which imply that there are only seven independent matrix elements.
This is consistent with the earlier result that there are at most seven parameters
$\tilde{g}',\beta_1,\beta_2,\beta_3,\alpha_1,\alpha_{24},\alpha_{25}$
 related to mixings. In Ref.\cite{Z'our}, we
had obtained a set of relations between matrix elements $U_{ij}$ and
parameters $g,g',g^{\prime\prime},\tilde{g}'$,
$\beta_1,\beta_2,\beta_3$,
 $\alpha_1,\alpha_8,\alpha_{24},\alpha_{25}$ as follows
\begin{eqnarray}
U\equiv\left(\begin{array}{ccc}
    \frac{1}{2g}c_\alpha
        &\frac{1}{2g}
        &-\frac{1}{2g}s_\alpha
    \\
    -\frac{1}{2g'}c_\alpha
        &\frac{1}{2g'}
        &\frac{1}{2g'}s_\alpha
    \\
    \frac{1}{g^{\prime\prime}}(s_\alpha+\frac{\tilde{g}'}{2g'}c_\alpha)
    &-\frac{\tilde{g}'}{2g^{\prime\prime}g'}
        &\frac{1}{g^{\prime\prime}}(c_\alpha-\frac{\tilde{g}'}{2g'}s_\alpha)
\end{array}\right)
\left(\begin{array}{ccc}
    \frac{c_\beta}{A_1}
        &0
        &\frac{s_\beta}{A_1}
    \\
    ga
        &gb
        &gc
    \\
    -\frac{s_\beta}{A_2}
        &0
        &\frac{c_\beta}{A_2}
\end{array}\right)
\left(\begin{array}{ccc}
    \frac{M_Z}{f}&0&0\\
    0&1&0\\
    0&0&\frac{M_{Z'}}{f}
\end{array}
\right)\;, ~~~~~\label{Udef1}
\end{eqnarray}
where $c_\alpha\equiv\cos\alpha_{Z'}$, $s_\alpha\equiv\sin\alpha_{Z'}$, $s_\beta=\sin\beta_{Z'}$,
$c_\beta=\cos\beta_{Z'}$ as well as the following definitions
\begin{eqnarray}
A_1^2=\frac{1}{4}(1\!-\!2\beta_1)c_\alpha^2+\beta_2s_\alpha
c_\alpha+(1\!-\!2\beta_3)s_\alpha^2 \hspace{1cm}
A_2^2=\frac{1}{4}(1\!-\!2\beta_1)s_\alpha^2-\beta_2s_\alpha
c_\alpha+(1-2\beta_3)c_\alpha^2\;,~~~\\
\tan\alpha_{Z'}=\frac{3+2\beta_1-8\beta_3-\sqrt{
(3+2\beta_1-8\beta_3)^2 +16\beta_2^2}}{4\beta_2}\hspace{1cm}
\tan\beta_{Z'}=\frac{-G_2+\sqrt{G_2^2+4G_0^2}}{2G_0}~~~~~~\label{alphaZ'}
\end{eqnarray}
\begin{eqnarray}
a&=&\frac{1}{gA_1A_2[{g'}^2{g^{\prime\prime}}^2-{g}^2{g'}^2{g^{\prime\prime}}^2(2\alpha_1+\alpha_8)
    +g^2{g^{\prime\prime}}^2
    -4g^2g'{g^{\prime\prime}}^2\tilde{g}'(\alpha_{24}+\alpha_{25})+g^2\tilde{g}^{\prime 2}]}
    \nonumber\\
    &&\times\Big\{[g^2{g''}^2+g^2\tilde{g}^{\prime 2}-g'^2{g''}^2+g^2g'^2{g''}^2\alpha_8+4g^2g'{g''}^2\tilde{g}'\alpha_{25}]
    (s_\alpha s_\beta A_1+c_\alpha c_\beta A_2)\nonumber\\
    &&+[2g^2g'\tilde{g}'+4g^2g'^2{g''}^2(\alpha_{24}+\alpha_{25})](-c_\alpha s_\beta A_1+s_\alpha c_\beta
    A_2)\Big\}\;.\nonumber\\
b^2&=&\frac{4{g'}^2{g^{\prime\prime}}^2}{(g^2+{g'}^2){g^{\prime\prime}}^2+g^2\tilde{g}^{\prime2}-{g}^2{g'}^2{g^{\prime\prime}}^2(2\alpha_1+\alpha_8)
    +4g^2g'{g^{\prime\prime}}^2\tilde{g}'(\alpha_{24}+\alpha_{25})}\;.
\nonumber\\
c&=&\frac{1}{gA_1A_2[{g'}^2{g^{\prime\prime}}^2-{g}^2{g'}^2{g^{\prime\prime}}^2(2\alpha_1+\alpha_8)
    +g^2{g^{\prime\prime}}^2
    -4g^2g'{g^{\prime\prime}}^2\tilde{g}'(\alpha_{24}+\alpha_{25})+g^2\tilde{g}^{\prime2}]}
    \nonumber\\
    &&\times\Big\{[g^2{g''}^2+g^2\tilde{g}^{\prime2}-g'^2{g''}^2+g^2g'^2{g''}^2\alpha_8
    +4g^2g'{g''}^2\tilde{g}'\alpha_{25}](-s_\alpha c_\beta A_1+c_\alpha s_\beta A_2)\nonumber\\
    &&
    +[2g^2g'\tilde{g}'+4g^2g'^2{g''}^2(\alpha_{24}+\alpha_{25})](c_\alpha c_\beta A_1+s_\alpha s_\beta A_2)
    \Big\}\;.~~~~
\nonumber\\
G_0&=&-A_1A_2\Big\{(-g^2-g'^2+{g^{\prime\prime}}^2+(\tilde{g}')^2)c_\alpha
s_\alpha
    +g'\tilde{g}'(s_\alpha^2-c_\alpha^2)
    +g^2[2g'^2c_\alpha s_\alpha+g'\tilde{g}'(c_\alpha^2-s_\alpha^2)]\alpha_1
    \nonumber\\
    &&+g^2[(g'^2-{g^{\prime\prime}}^2-(\tilde{g}')^2)c_\alpha s_\alpha -g'\tilde{g}'(s_\alpha^2-c_\alpha^2)]\alpha_8
    +2g^2{g^{\prime\prime}}^2(c_\alpha^2-s_\alpha^2)(\alpha_{24}+g'^2\alpha_1\alpha_{25})
    \nonumber\\
    &&+{g^{\prime\prime}}^2[-4g'\tilde{g}'c_\alpha s_\alpha+2g'^2(c_\alpha^2-s_\alpha^2)]
        [g^2(\alpha_8\alpha_{25}-\alpha_1\alpha_{24})-\alpha_{25}]+g^2{g^{\prime\prime}}^2[8g'^2s_\alpha c_\alpha
    \nonumber\\
    &&+4g'\tilde{g}'(c_\alpha^2-s_\alpha^2)]\alpha_{24}\alpha_{25}
    +g^2g'^2{g^{\prime\prime}}^2s_\alpha c_\alpha(4\alpha_{25}^2-\alpha_1^2)
   +4g^2{g^{\prime\prime}}^2(g's_\alpha+\tilde{g}'c_\alpha)(g'c_\alpha-\tilde{g}'s_\alpha)\alpha_{24}^2
    \Big\}
\nonumber\\
G_2&=&A_1^2\Big\{(g^2+g'^2)c_\alpha^2
    +({g^{\prime\prime}}^2+(\tilde{g}')^2)s_\alpha^2(1-g^2\alpha_8)
    -g^2g'^2c_\alpha^2(2\alpha_1+\alpha_8)
    +4g'{g^{\prime\prime}}^2\tilde{g}'s_\alpha^2\alpha_{25}
    \nonumber\\
    &&-4g^2g'^2{g^{\prime\prime}}^2c_\alpha^2(\alpha_{24}^2+\alpha_{25}^2+2\alpha_{24}\alpha_{25})
    -g^2{g^{\prime\prime}}^2s_\alpha^2[g'^2\alpha_1^2+4(\tilde{g}')^2\alpha_{24}^2+4g'\tilde{g}'(\alpha_8\alpha_{25}-\alpha_1\alpha_{24})]
    \Big\}
    \nonumber\\
    &&-[A_1\rightarrow A_2,c_\alpha\leftrightarrow s_\alpha]
    +s_\alpha c_\alpha(A_1^2+A_2^2)\Big\{-2g'\tilde{g}'[1-g^2(\alpha_1+\alpha_8)]
    \nonumber\\
    &&+4g^2{g^{\prime\prime}}^2[(\alpha_{24}-\alpha_{25})(1-{g''}^2\alpha_1)+2g'\tilde{g}'\alpha_{24}^2+{g''}^2\alpha_8\alpha_{25}]\Big\}\;.
    \label{G2}
\end{eqnarray}
Finally the masses of $Z$ and $Z'$ bosons are determined from
\begin{eqnarray}
\mathbf{K}_{11}=-\frac{1}{4}\hspace{1cm}
\mathbf{K}_{33}=-\frac{1}{4}\;,
\end{eqnarray}
with
\begin{eqnarray}
\mathbf{K}\equiv U^T\left(\begin{array}{ccc}
    -\frac{1}{4}(1-\alpha_8g^2)&\frac{1}{4}\alpha_1 gg'&\frac{1}{2}gg^{\prime\prime}\alpha_{24}\\
    \frac{1}{4}\alpha_1 gg'&-\frac{1}{4}&\frac{1}{2}g'g^{\prime\prime}\alpha_{25}\\
    \frac{1}{2}gg^{\prime\prime}\alpha_{24}&\frac{1}{2}g'g^{\prime\prime}\alpha_{25}&-\frac{1}{4}\\
\end{array}\right)U\;.~~~~~~~~
\end{eqnarray}
General expressions for the mixing matrix elements $U_{ij}$ are too
complicated to be written analytically. In Ref.\cite{Z'our}, we listed
results for $U_{ij}$, $M_Z$ and $M_{Z'}$  expanded up to order
$p^4$ and linear in $\tilde{g}'$. In real new physics models
appearing in the literature, the $Z'\!-\!Z\!-\!\gamma$ mixings are
often not so complex. In the next section, we identify and discuss
typical $Z'\!-\!Z\!-\!\gamma$ mixings appearing in various new physics models.
\section{Classification of models in terms of their $Z'\!-\!Z\!-\!\gamma$ mixings}\label{MixingModels}

In this section, we organize the various new physics models that can be
found in the literature involving the $Z'$ boson according to their $Z'\!-\!Z\!-\!\gamma$ mixings.
Unlike the most general case reviewed in the last section, these mixings are special
$Z'\!-\!Z\!-\!\gamma$ mixings for which the mixing matrix elements $U_{ij}$ and $M_Z$, $M_{Z'}$
can all be work out exactly. Below we consider five situations.
\begin{enumerate}
\item \underline{\bf Minimal $Z'\!-\!Z$ mass mixing}~\cite{RizzoARXIV2006,FranziniPRD1987,RizzoPRD1991,LangackerPRD1992,ChiappettaPRD1996,
FramptonPRD1996,ErlerPRL2000,AnokaNPB2004,KozlovPRD2005,BassoARXIV2008,ChanowitzARXIV2008,AppelquistPRD2003,
FerrogliaAP2007,CarenaPRD2004}:\\
This kind of model provides minimal
mixing by ignoring all mixings
in the kinetic terms and $Z\!-\!\gamma$,~$Z'\!-\!\gamma$ mixings in
the mass terms. They correspond to the vanishing five parameters
\begin{eqnarray}
\tilde{g}'=\alpha_1=\alpha_8=\alpha_{24}=\alpha_{25}=0\;.
\end{eqnarray}
With the exception of gauge couplings $g,g',g^{\prime\prime}$, the remaining three
nontrivial parameters are denoted by the $Z'\!-\!Z$ mass matrix
 \begin{eqnarray}
        \mathcal{M}^2=\left(\begin{array}{cc}
            M_{Z_0}^2&M_{ZZ'}^2\\
            M_{ZZ'}^2& M_{Z'_0}^2\end{array}\right)
            \hspace{1cm}Z_0^\mu\equiv\frac{gW^3_\mu\!-g'B_\mu}{\sqrt{g^2\!+g^{\prime2}}}\hspace{0.5cm}
            A_0^\mu\equiv\frac{g'W^3_\mu\!+gB_\mu}{\sqrt{g^2\!+g^{\prime2}}}\hspace{0.5cm}
            Z_0^{\prime\mu}\equiv X^\mu,~~~~
        \label{MinimalZZpmassmixing}
        \end{eqnarray}
where mass parameters $M_{Z_0}^2$, $M_{Z'_0}^2$ and $M_{ZZ'}^2$
are related to $\beta_1,\beta_2,\beta_3$ as
\begin{eqnarray}
\frac{f^2}{4}(1\!-\!2\beta_1)(g^2\!+\!g^{\prime2})\equiv
M_{Z_0}^2\hspace{0.6cm}f^2(1\!-\!2\beta_3)g^{\prime\prime2}\equiv
M_{Z_0'}^2\hspace{0.6cm}\frac{f^2}{2}\beta_2g^{\prime\prime}\sqrt{g^2\!+\!g^{\prime2}}\equiv
M_{ZZ'}^2\;.~~~~~
\end{eqnarray}
Refs.\cite{AppelquistPRD2003,FerrogliaAP2007} use an alternative
expression which corresponds to setting
\begin{eqnarray}
f=v_H\hspace{0.5cm}g'=g_Y\hspace{0.5cm}g^{\prime\prime}=g_z\hspace{0.5cm}\beta_1=0\hspace{0.5cm}
\beta_2=-\frac{1}{2}z_H\hspace{0.5cm}
1-2\beta_3=\frac{1}{4}(z_H^2+\frac{v_{\phi}^2}{f^2})\;.~~~~~~~~~\nonumber
\end{eqnarray}
Ref.\cite{CarenaPRD2004} further generalizes this which leads then to
\begin{eqnarray}
g'\!=g_Y\hspace{0.4cm} g^{\prime\prime}\!\!=g_z\hspace{0.4cm}
1-2\beta_1=\frac{v_{H_1}^2+v_{H_2}^2}{f^2}~~~
\beta_2\!=-\frac{z_{H_1}v_{H_1}^2\!\!+\!z_{H_2}v_{H_2}^2}{2f^2}\hspace{0.4cm}
\nonumber\\
1\!-\!2\beta_3\!=\frac{1}{4f^2}(z_{H_1}^2v_{H_1}^2\!\!+\!z_{H_2}^2v_{H_2}^2\!\!+\!v_{\phi}^2)\;.~~~~~~~~~\nonumber
\end{eqnarray}
In this kind of model, the key $Z'\!-\!Z$ mixing parameter is
$\beta_2$
 which yields a  non-vanishing off-diagonal element $M_{ZZ'}^2$ in the
$Z'\!-\!Z$ mass matrix. This element further generates the seesaw
splitting between the original $Z$ and $Z'$ masses,
\begin{eqnarray}
M_Z^2&=&\frac{1}{2}[M_{Z_0}^2\!+M_{Z_0'}^2\!-\sqrt{(M_{Z_0}^2\!-M_{Z_0'}^2)^2\!+4M_{ZZ'}^4}]\approx
M_{Z_0}^2-\frac{M_{ZZ'}^4}{M_{Z_0'}^2\!-M_{Z_0}^2}\\
M_{Z'}^2&=&\frac{1}{2}[M_{Z_0}^2\!+M_{Z_0'}^2\!+\sqrt{(M_{Z_0}^2\!-M_{Z_0'}^2)^2\!+4M_{ZZ'}^4}]
\approx M_{Z_0'}^2+\frac{M_{ZZ'}^4}{M_{Z_0'}^2\!-M_{Z_0}^2}\;.~~~~
\end{eqnarray}
Meanwhile the $Z'\!-\!Z$ mixing can be parameterized by mixing angle $\theta'$
\begin{eqnarray}
  \left(\begin{array}{c}Z_0^\mu\\Z_0^{\prime\mu}\end{array}\right)=\left(\begin{array}{cc}
            \cos\theta'&\sin\theta'\\
            -\sin\theta'&\cos\theta'\end{array}\right)\left(\begin{array}{c}Z^\mu\\Z^{\prime\mu}\end{array}\right)
            \hspace{1cm}\tan
            2\theta'=\frac{2M_{ZZ'}^2}{M_{Z_0'}^2\!-M_{Z_0}^2}\;.~~~~\label{thetapdef}
\end{eqnarray}
leading to a rotation matrix introduced in (\ref{Udef}) of the form
\begin{eqnarray}
U_{\mbox{\tiny Minimal $Z'\!\!\!-\!\!Z$ mass mixing}}&=&
    \left(\begin{array}{ccc}\cos\theta_W&\sin\theta_W&0\\-\sin\theta_W&\cos\theta_W&0\\0&0&1\end{array}\right)
    \left(\begin{array}{ccc}\cos\theta'&0&\sin\theta'\\ 0&1&0\\ -\sin\theta'&0&\cos\theta'\end{array}\right)
    \nonumber\\
&=&\left(\begin{array}{ccc}\cos\theta_W\cos\theta'&\sin\theta_W&\cos\theta_W\sin\theta'\\
        -\sin\theta_W\cos\theta&\cos\theta_W&-\sin\theta_W\sin\theta'\\
        -\sin\theta'&0&\cos\theta'
        \end{array}\right)\;,
\end{eqnarray}
with an electroweak mixing angle $\tan\theta_W=g'/g$.
\item \underline{\bf Minimal $Z'\!-\!Z$ kinetic mixing}~\cite{RizzoARXIV1998,LangackerPRD2008,LangackerPRL2008}:\\
This kind of model provides minimal mixing by ignoring all mixings
in the mass terms and $Z\!-\!\gamma$,~$Z'\!-\!\gamma$ mixings in the
kinetic terms leading to the vanishing of seven parameters
\begin{eqnarray}
\tilde{g}'=\beta_1=\beta_2=\beta_3=\alpha_1=\alpha_8=\alpha_{24}=0\;.
\end{eqnarray}
Again with the exception of gauge couplings $g,g',g^{\prime\prime}$, the one remaining
nontrivial parameter is denoted by
\begin{eqnarray}
 g'g^{\prime\prime}\alpha_{25}\equiv-\frac{\sin\chi}{2}\;.
\end{eqnarray}
following Ref.\cite{RizzoARXIV1998}, we redefine the gauge fields as
\begin{eqnarray}
B^\mu=B_0^\mu-\tan\chi
Z^{\prime\mu}_0\hspace{2cm}X^\mu=\frac{Z^{\prime\mu}_0}{\cos\chi}\;,\label{Bredef}
\end{eqnarray}
in terms of the fields $B_0^\mu,Z_0^{\prime\mu},W^{3\mu}$, the
kinetic term appears diagonalized and the model reduces to a minimal
$Z'-Z$ mass mixing model discussed above\footnote{This detail was not pointed out in Ref.\cite{RizzoARXIV1998}.} with
\begin{eqnarray}
M_{Z_0}^2=\frac{f^2}{4}(g^2\!+\!g^{\prime2})
\hspace{0.6cm}M_{Z_0'}^2=\frac{f^2[g^{\prime2}\sin^2\chi\!+\!4g^{\prime\prime2}]}{4\cos^2\chi}
\hspace{0.6cm}M_{ZZ'}^2=\frac{f^2}{4}g'\sqrt{g^2\!+\!g^{\prime2}}\tan\chi
\;.~~~~~
\end{eqnarray}
The rotation matrix introduced in (\ref{Udef}) takes the form
\begin{eqnarray}&&\hspace{-1cm}U_{\mbox{\tiny Minimal $Z'\!\!\!-\!\!Z$ kinetic mixing}}=
\left(\begin{array}{ccc}1&0&0\\0&1&-\tan\chi\\0&0&\frac{1}{\cos\chi}
\end{array}\right)\times U_{\mbox{\tiny Minimal $Z'\!\!\!-\!\!Z$ mass mixing}}\nonumber\\
&&\hspace{-1cm}= \left(\begin{array}{ccc}
    \cos\theta'\cos\theta_W&\sin\theta_W&\cos\theta_W\sin\theta'\\
    -\sin\theta_W\cos\theta'+\tan\chi\sin\theta'&\cos\theta_W&-\sin\theta_W\sin\theta'-\tan\chi\cos\theta'\\
    -\sin\theta'/\cos\chi&0&\cos\theta'/\cos\chi
\end{array}\right)
\;.~~~~~\label{KineticU0}
\end{eqnarray}
\item \underline{\bf General $Z'\!-\!Z$ mixing}~\cite{RizzoARXIV2006,Hill,Lane,Chiv,CasselARXIV2009,Holdom1986,PDG2006,BabuPRD1996}:\\
This kind of model
is combinations of minimal $Z'\!-\!Z$ mass
mixing model and minimal $Z'\!-\!Z$ kinetic mixing model
discussed above which correspond to
\begin{eqnarray}
\tilde{g}'=\alpha_1=\alpha_8=\alpha_{24}=0\hspace{1cm}
g'g^{\prime\prime}\alpha_{25}\equiv-\frac{\sin\chi}{2}\;.
\end{eqnarray}
In a similar manner as for minimal $Z'\!-\!Z$ kinetic mixing model, we can use
(\ref{Bredef}) to remove the mixing in the kinetic term and then, in
terms of the fields $B_0^\mu,Z_0^{\prime\mu},W^{3\mu}$, the model can be
changed into a minimal $Z'\!-\!Z$ mass mixing model with identifications
\begin{eqnarray}
M_{Z_0}^2&=&\frac{f^2}{4}(1-2\beta_1)(g^2\!+g^{\prime2})\nonumber\\
M_{Z_0'}^2&=&\frac{f^2[g^{\prime2}(1-2\beta_1)\sin^2\chi+4g^{\prime\prime2}(1-2\beta_3)
+4\beta_2g'g^{\prime\prime}\sin\chi]}{4\cos^2\chi}\nonumber\\
M_{ZZ'}^2&=&\frac{f^2}{4}\frac{(1-2\beta_1)g'\sin\chi+2\beta_2g^{\prime\prime}}{\cos\chi}\sqrt{g^2\!+g^{\prime2}}
\;.~~~~~\label{M000}
\end{eqnarray}
The resulting rotation matrix has the same form as in (\ref{KineticU0}),
the only change is that now the $\theta'$ as determined through
(\ref{thetapdef}) is different due to the new expressions for
$M_{Z_0}^2,M_{Z_0'}^2,M_{ZZ'}^2$ given by (\ref{M000}).
In some dynamical models such as TC2 models, the general $Z'\!-\!Z$ mixings
are generated by technicolor and topcolor dynamics, as in
Refs.\cite{Hill1,Lane1,Chiv1},
while mixing parameters are given through
dynamical computations depending on the nature of the
TC2 models
 and results
in the following expressions
\begin{eqnarray}
g'g^{\prime\prime}\alpha_{25}=\frac{g^{\prime2}\gamma}{2c_{Z'}}\hspace{2cm}
\frac{f^2}{2}\beta_2g^{\prime\prime}=\frac{g'}{4c_{Z'}}\times\left\{\begin{array}{lll}
(F_0^{\mathrm{TC2}})^2\tan\theta'&~~~&\mbox{Ref.\cite{Hill,Hill1}}\\
3(F_0^{\mathrm{1D}})^2\tan\theta'&~~~&\mbox{Ref.\cite{Lane,Lane1}}\\
-3(F_0^{\mathrm{1D}})^2\cot\theta'&~~~&\mbox{Ref.\cite{Chiv,Chiv1}}\end{array}\right.\;,
\end{eqnarray}
where all symbols appearing on the right-hand side of these results
are parameters pertaining to the TC2 models.
\item \underline{\bf $Z'\!-\!\gamma$ kinetic and $Z'\!-\!Z$ mixing}~\cite{HoldomPLB1991}:\\
    B. Holdom extends the conventional $Z'\!-\!Z$ mixing by further adding in
    model a $Z'\!-\!\gamma$ kinetic mixing term. His model
    corresponds to having
\begin{eqnarray}
&&\hspace{-0.5cm}\tilde{g}'=\alpha_1=\alpha_8=0\hspace{0.5cm}\frac{f^2}{4}(1\!-\!2\beta_1)=m_Z^2\hspace{0.5cm}
\frac{f^2}{2}\beta_2g^{\prime\prime}
\sqrt{g^2\!+g^{\prime2}}=xm_Z^2\hspace{0.5cm}f^2(1\!-\!2\beta_3)g^{\prime\prime2}=m_X^2\nonumber\\
&&\hspace{-0.5cm}gg^{\prime\prime}\sqrt{g^2\!+g^{\prime2}}\alpha_{24}=-\frac{1}{2}(gy+g'w)\hspace{1cm}
gg^{\prime\prime}\sqrt{g^2\!+g^{\prime2}}\alpha_{25}=\frac{1}{2}(g'y-gw)\;.
\end{eqnarray}
We can diagonalize the kinetic terms by redefining the $B^\mu$ and
$W^{3\mu}$ fields as
\begin{eqnarray}
&&\hspace{-0.5cm}B^\mu=B_0^\mu-\frac{\sin\chi}{\sqrt{1-\sin^2\chi-\sin^2\overline{\chi}}}Z_0^{\prime\mu}\hspace{1cm}
W^{3\mu}=W^{3\mu}_0-\frac{\sin\overline{\chi}}{\sqrt{1-\sin^2\chi-\sin^2\overline{\chi}}}
Z_0^{\prime\mu}
~~~~~~~\\
&&\hspace{-0.5cm}X^\mu=\frac{Z_0^{\prime\mu}}{\sqrt{1-\sin^2\chi-\sin^2\overline{\chi}}}\hspace{1cm}
-\frac{\sin\overline{\chi}}{2}\equiv
g'g^{\prime\prime}\alpha_{24}\hspace{1cm}-\frac{\sin\chi}{2}\equiv
g'g^{\prime\prime}\alpha_{25}\nonumber
\end{eqnarray}
and then in terms of fields $B_0^\mu,Z_0^{\prime\mu},W^{3\mu}_0$,
the model becomes the minimal $Z'\!-\!Z$ mass mixing model with
\begin{eqnarray}
M_{Z_0}^2&=&\frac{f^2}{4}(1-2\beta_1)(g^2\!+g^{\prime2})\nonumber\\
M_{Z_0'}^2&=&\frac{f^2[\frac{1}{4}(1-2\beta_1)(g'\sin\chi-g\sin\overline{\chi})^2+(1-2\beta_3)g^{\prime\prime2}+
\beta_2g^{\prime\prime}(g'\sin\chi-g\sin\overline{\chi})]}{1-\sin^2\chi-\sin^2\overline{\chi}}\nonumber\\
M_{ZZ'}^2&=&\frac{f^2}{4}\frac{[(1-2\beta_1)(g'\sin\chi-g\sin\overline{\chi})+2\beta_2g^{\prime\prime}]}
{\sqrt{1-\sin^2\chi-\sin^2\overline{\chi}}}\sqrt{g^2\!+g^{\prime2}}
\;.~~~~~\label{M1}
\end{eqnarray}
for which the rotation matrix introduced in (\ref{Udef}) takes the form
\begin{eqnarray}&&\hspace{-1cm}U_{\mbox{\tiny $Z'\!\!\!-\!\!\gamma$ kinetic and $Z'\!\!\!-\!\!Z$ mixing}}=
\left(\begin{array}{ccc}1&0&-\frac{\sin\overline{\chi}}{\sqrt{1-\sin^2\chi-\sin^2\overline{\chi}}}\\
0&1&-\frac{\sin\chi}{\sqrt{1-\sin^2\chi-\sin^2\overline{\chi}}}\\0&0&\frac{1}{\sqrt{1-\sin^2\chi-\sin^2\overline{\chi}}}
\end{array}\right)\times U_{\mbox{\tiny Minimal $Z'\!\!\!-\!\!Z$ mass mixing}}\nonumber\\
&&\hspace{-1cm}=\left(\begin{array}{ccc}\frac{g\cos\theta'}{\sqrt{g^2+g^{\prime2}}}+
\frac{\sin\theta'\sin\overline{\chi}}{\sqrt{1-\sin^2\chi-\sin^2\overline{\chi}}}&~~~\frac{g'}{\sqrt{g^2+g^{\prime2}}}~~~&
\frac{g\sin\theta'}{\sqrt{g^2+g^{\prime2}}}-\frac{\cos\theta'\sin\overline{\chi}}{\sqrt{1-\sin^2\chi-\sin^2\overline{\chi}}}\\
-\frac{g'\cos\theta'}{\sqrt{g^2+g^{\prime2}}}+\frac{\sin\theta'\sin\chi}{\sqrt{1-\sin^2\chi-\sin^2\overline{\chi}}}
&\frac{g}{\sqrt{g^2+g^{\prime2}}}&
-\frac{g'\sin\theta'}{\sqrt{g^2+g^{\prime2}}}-\frac{\cos\theta'\sin\chi}{\sqrt{1-\sin^2\chi-\sin^2\overline{\chi}}}\\
-\frac{\sin\theta'}{\sqrt{1-\sin^2\chi-\sin^2\overline{\chi}}}&0&\frac{\cos\theta'}{\sqrt{1-\sin^2\chi-\sin^2\overline{\chi}}}
\end{array}\right)\;.~~~~~\label{KineticU}
\end{eqnarray}
\item \underline{\bf Stueckelberg-type mixing}~\cite{KorsJHEP2005,FeldmanPRL2006,FeldmanPRD2007}:\\
This kind of model provides mixing through the nonzero coupling
constant $\tilde{g}'$ and except for gauge coupling $g,g',g^{\prime\prime}$,
a typical choice as given in Refs.\cite{KorsJHEP2005,FeldmanPRL2006} is
the vanishing of all other parameters
\begin{eqnarray}
\beta_1=\beta_2=\beta_3=\alpha_1=\alpha_8=\alpha_{24}=\alpha_{25}=0\;,
\end{eqnarray}
leading to diagonal kinetic terms and mixing occurring only in
the mass terms. After rotating the standard electroweak mixing angle
$\theta_W$, we can redefine the gauge fields
\begin{eqnarray}
\bar{B}^\mu&=&-\frac{{g''}\sqrt{g^2+g'^2}}{(g^2+g'^2){g''}^2+g^2\tilde{g}'^2}B_{0}^\mu
        +\frac{g\tilde{g}'}{(g^2+g'^2){g''}^2+g^2\tilde{g}'^2}{Z'}_0^\mu
    \nonumber\\
\bar{Z}'^\mu&=&\frac{g\tilde{g}'}{(g^2+g'^2){g''}^2+g^2\tilde{g}'^2}B_0^\mu
        +\frac{{g''}\sqrt{g^2+g'^2}}{(g^2+g'^2){g''}^2+g^2\tilde{g}'^2}{Z'}_0^\mu
\label{Stueck}\end{eqnarray}
thereby changing the present model to a minimal
$Z'-Z$ mass mixing model with
\begin{eqnarray}
M_{Z_0}^2&=&\frac{f^2}{4}(g^2+g'^2+\frac{4g'^2\tilde{g}'^2}{g^2+g'^2})
\nonumber\\
M_{Z'_0}^2&=&f^2({g''}^2+\frac{g^2\tilde{g}'^2}{g^2+g'^2})
\nonumber\\
M_{ZZ'}&=&-\frac{f^2g'\tilde{g}'\sqrt{(g^2+g'^2){g''}^2+g^2\tilde{g}'^2}}{g^2+g'^2}\;.
\label{StueckZZprimemass}
\end{eqnarray}
The overall rotation matrix then becomes
\begin{eqnarray}
U_{\mbox{\tiny Stuekckelberg type mixing}}&=&
\left(\begin{array}{ccc}\cos\theta_W&\sin\theta_W&0\\-\sin\theta_W&\cos\theta_W&0\\0&0&1\end{array}\right)
    \left(\begin{array}{ccc}
        1&0&0\\
        0&-\frac{{g''}\sqrt{g^2+g'^2}}{(g^2+g'^2){g''}^2+g^2\tilde{g}'^2}&\frac{g\tilde{g}'}{(g^2+g'^2){g''}^2+g^2\tilde{g}'^2}\\
        0&\frac{g\tilde{g}'}{(g^2+g'^2){g''}^2+g^2\tilde{g}'^2}&\frac{{g''}\sqrt{g^2+g'^2}}{(g^2+g'^2){g''}^2+g^2\tilde{g}'^2}\end{array}\right)
    \nonumber\\
    &&\times\left(\begin{array}{ccc}\cos\theta'&0&\sin\theta'\\ 0&1&0\\ -\sin\theta'&0&\cos\theta'\end{array}\right)
\end{eqnarray}
with $\theta'$ evaluated from the second equation of
(\ref{thetapdef}) and those of (\ref{StueckZZprimemass}). In
Ref.\cite{FeldmanPRD2007}, the Stueckelberg-type mixing is further
generalized to include kinetic mixing by relaxing the original
condition $\alpha_{25}=0$. This kinetic mixing can be diagonalized
by applying (\ref{Bredef}) and following a similar procedure to that leading to
(\ref{Stueck}) in diagonalizing the mass terms.
\end{enumerate}
\section{the $Z'$ boson charges to quark and leptons}\label{SECcharge}

The charges for the $Z'$ boson with respect to ordinary quarks and leptons can be
expressed in terms of the gauge interaction as
\begin{eqnarray}
\mathcal{L}_{\mathrm{gauge~coupling}}={g''}X_\mu J^\mu_X
\hspace{2cm}J^\mu_X=\sum_i\bar{f}_i\gamma^\mu[y'_{iL}P_L+y'_{iR}P_R]f_i\;,\label{NCgauge}
\end{eqnarray}
where index $i$ distinguishes the three generations associated with the six quarks
$u,c,t,d,s,b$ and six leptons $e,\mu,\tau,\nu_e,\nu_\mu,\nu_\tau$, and
$y_{i,L}',y_{i,R}'$ are the corresponding left- and right-hand
charges\footnote{Phenomenologically, we need to further express the gauge
interaction given in Eq.(\ref{NCgauge}) in terms of  mass
eigenstate of  $Z'$, for then the $Z'\!-\!Z\!-\!\gamma$ mixings discussed in
the last section set in.}. The $SU(2)_L$ symmetry requires equating $U(1)$
charges of the two components of the left-hand fermion doublet, i.e.
$y'_{u,L}=y'_{d,L}\equiv y'_q$ for quark and
$y'_{\nu,L}=y'_{e,L}\equiv y'_l$ for lepton. Thus, we can
parameterize the fermionic $U(1)'$ charges by $y'_q$, $y'_u$, $y'_d$,
$y'_l$, $y'_e$ and $y'_\nu$. In general, the assignments of $U(1)'$
charges are generation-dependent, but in its simplest form $U(1)'$
charges can be generation-independent, much like hypercharge assignments
in SM. TABLE.\ref{ZprimeCharge} lists four sets of more common assignments
for the generation-independent $U(1)'$ charges of fermions
in new physics models involving $Z'$ boson
\cite{CarenaPRD2004,PDG2008}. In the $U(1)_{B-xL}$ model (see column 3 of TABLE.\ref{ZprimeCharge}),
$Z'$ charges
are determined by the baryon number and lepton number from $y'_i=B_i-xL_i$
with a free rational parameter $x$. Leptophobic and hadrophobic $Z'$
models correspond to $x=\infty$ and $x=0$, respectively. The
second set of charges comes from grand unified theories. Parameter $x$
establishes the mixing of the two extra $U(1)$ groups in the $E_6$
symmetry breaking patterns $E_6\rightarrow SU(5)\times U(1)\times U(1)$.
$Z_\chi$, $Z_\psi$ and $Z_\eta$ of Ref.\cite{GUTs} correspond to the special case with
$x=-3$, $x=1$ and $x=-1/2$, respectively. The third set,
$U(1)_{d-xu}$ results in the vanishing of the left-hand quark doublet
charge and the ratio of right-hand up quark charges to down quark charges is controlled by
$-x$. In the last set, the free parameter $x$ is the ratio of the charges of the left-hand
quark doublet and right-hand up quark singlet and reduces to the $U(1)_{B-L}$ model for $x=1$.
\begin{table}
  \centering
  \caption{generation-independent $U(1)'$ charges for quarks and leptons}\label{ZprimeCharge}
\begin{tabular}{c|c|c|c|c|c}
  \hline\hline
  models & $Z'$ EWCL & $U(1)_{B-xL}$ & $U(1)_{10+x\bar{5}}$ & $U(1)_{d-xu}$ & $U(1)_{q+xu}$ \\
  \hline
  $(u_L, d_L)$   & $y'_{q}$   & $1/3$ & $1/3$  & $0$      &  $1/3$\\
  $u_R$          & $y'_{u}$   & $1/3$ & $-1/3$ & $-x/3$   &  $x/3$\\
  $d_R$          & $y'_{d}$   & $1/3$ & $-x/3$ & $1/3$    &  $(2-x)/3$\\
  \hline
  $(\nu_L, e_L)$ & $y'_{l}$   & $-x$  & $x/3$  & $(x-1)/3$&  $-1$\\
  $e_R$          & $y'_{e}$   & $-x$  & $-1/3$ & $x/3$    &  $-(2+x)/3$\\
  \hline
  $\nu_R$        &$y'_{\nu_R}$& $-1$   & $(x-2)/3$    & $-x/3$      &  $(x-4)/3$\\
  \hline\hline
\end{tabular}
\end{table}
 Theoretically, the charges of quarks and leptons must satisfy the anomaly cancellation conditions to
preserve the gauge symmetry. We now examine the constraints on
generation-independent $U(1)'$ charges arising as a consequence of these anomaly cancellation
conditions. Davidson et.al. \cite{DavidsonPRD1979} have studied
anomaly cancellation for additional $U(1)'$ gauge group and
derived the following anomaly cancellation conditions for $U(1)_Y\otimes
U(1)'$ gauge groups
\begin{eqnarray}
\sum y^\alpha_L=\sum
Q^2(y^\alpha_L\!-\!y^\alpha_R)=0\hspace{0.6cm}\sum
Q(y^\alpha_Ly^\beta_L\!-\!y^\alpha_Ry^\beta_R)=0\hspace{0.6cm}\sum
(y^\alpha_Ly^\beta_Ly^\gamma_L\!-\!y^\alpha_Ry^\beta_Ry^\gamma_R)=0\;,~~~~
\end{eqnarray}
where $\alpha,\beta,\gamma$
indexes $U(1)_Y$ and $U(1)'$
charges. Substituting the $U(1)_Y$ charges for ordinary quarks and
leptons and assuming the generation-independence of $U(1)'$
charges, we find that above equations imply
\begin{eqnarray}
\left\{\begin{array}{l} y'_l+3y'_q=0
\\
3y'_l+5y'_q-3y'_e-4y'_u-y'_d=0
\\
-{y'_l}^2+{y'_q}^2+{y'_e}^2-2{y'_u}^2+{y'_d}^2=0
\\
3y'_l+y'_q-6y'_e-8y'_u-2y'_d=0
\\
2{y'_l}^3+6{y'_q}^3-{y'_e}^3-3{y'_u}^3-3{y'_d}^3-{y'_{\nu_R}}^3=0
\end{array}\right.\;.\label{constraint}
\end{eqnarray}
The last equation in (\ref{constraint}) can be satisfied by
assigning $y'_{\nu_R}$ a proper value or adding in our theory some
other new fermions. Solving the above equations, we obtain two sets of solutions
which satisfy the anomaly cancellation conditions
\begin{eqnarray}
\left\{\begin{array}{l}
    y'_l=-3y'_q\\
    y'_d=2y'_q-y'_u\\
    y'_e=-2y'_q-y'_u\\
    y'_{\nu_R}=-4y'_q+y'_u
\end{array}\right.
\hspace{2cm}{\rm or}\hspace{2cm}\left\{\begin{array}{l}
    y'_l=-3y'_q\\
    y'_d=-\frac{14}{5}y'_q+\frac{1}{5}y'_u\\
    y'_e=-\frac{2}{5}y'_q-\frac{7}{5}y'_u\\
    y'_{\nu_R}=\frac{\sqrt[3]{35}}{5}(4y'_q-y'_u)
\end{array}\right.\;.\label{TwoSolutions}
\end{eqnarray}
Of the six of $U(1)'$ charges,  only two of them $y'_q$ and
$y'_u$ are independent; the other four being linear
combinations of these two. In addition, there are two kinds of linear
combinations: the first of Eq.(\ref{TwoSolutions}) which was given and discussed in detail in
Ref.\cite{AppelquistPRD2003}, while the second is a new solution having not yet appeared in the literature.
We can utilize the values of $y'_q$ and
$y'_u$ to classify the new physics models and in the following we
list some typical cases:
\begin{enumerate}
\item Left Handed:~$y'_u=y'_d=y'_e=y'_{\nu_R}=0~\Rightarrow~y'_q=y'_l=0$
\item Right
Handed:~$y'_q=y'_l=0~\Rightarrow~y'_d\!=-y'_u\!=y'_e\!=-y'_{\nu_R}$
or
$y'_d\!=\frac{1}{5}y'_u\!=-\frac{1}{7}y'_e\!=-\frac{1}{\sqrt[3]{35}}y'_{\nu_R}$
\item Left-Right symmetric:~$y'_q=y'_u=y'_d~\Rightarrow~y'_l=y'_e=y'_{\nu_R}=-3y'_q$
\item $\nu_R$
decouple:~$y'_{\nu_R}=0~\Rightarrow~y'_u=4y'_q,~y'_e=2y'_l=3y'_d=-6y'_q$
\end{enumerate}
Checking the
assignments given in TABLE.\ref{ZprimeCharge} against the two solutions in (\ref{TwoSolutions}),
we find that the
$U(1)_{B-xL}$, $U(1)_{d-xu}$ and $U(1)_{q+xu}$ models are anomaly-free when parameter $x=1$
with the right-hand neutrino charge
$y'_{\nu_R}=-1$, $y'_{\nu_R}=-\frac{1}{3}$ and $y'_{\nu_R}=-1$,
respectively. Furthermore, the $U(1)_{10+x\bar{5}}$ model is anomaly-free when
$x=-3$ with $y'_{\nu_R}=-5/3$. Even though the anomaly cancellation
condition can not be satisfied with the present quarks and leptons,
we still have the possibility of canceling the anomalies by
adding some extra fermions into theory.

If we relax the generation-independence criterion on the $U(1)'$
charges, we need to add generation indices to each of the charges in
Eq.(\ref{constraint}) and sum over the generations on the left-hand
side of Eq.(\ref{constraint}). In this case, there are too many free
parameters and solutions. We list
 several possible solutions in TABLE.\ref{GenDependentCharge}, in which the first and last columns
are the two solutions given in Ref.\cite{PDG2008},
and the remaining solutions can
be seen to be some kind
 of generation-dependent generalization of charge assignments
given in the third, fourth
and fifth columns in TABLE.\ref{ZprimeCharge}. The typical feature
of these solutions is that for the solutions given in the first
four columns of TABLE.\ref{GenDependentCharge}, the charges
for the first two generations are parameterized in a like manner as those in
the generation-independent situation by $x$ or $y$ separately, and differences appear only
in the third generation of quarks and leptons.
Of special note is that for the solution to $U(1)_{q+xu+yc+zt}$, the anomaly
cancellation condition is satisfied for each generation
independently.
\begin{table}
  \centering
  \caption{generation-dependent charge}\label{GenDependentCharge}
\begin{tabular}{c|c|c|c|c|c}
  \hline\hline
  models & $U(1)_{B-xL_e\!-yL_\mu\!}$ & $U(1)_{10+x\bar{5}}~
  \mathrm{\tiny gen\!\!-\!dep}$ & $U(1)_{d-xu}~\mathrm{\tiny gen\!\!-\!dep}$ & $U(1)_{q+xu+yc+zt}$ & $2\!+\!1~
  \mathrm{\tiny leptocratic}$\\
  \hline
  $q_{1,L}$     & $1/3$ & $1/3$  & $0$      &  $1/3$ & $1/3$\\
  $u_R$         & $1/3$ & $-1/3$ & $-x/3$   &  $x/3$ & $x/3$\\
  $d_R$         & $1/3$ & $-x/3$ & $1/3$    &  $(2-x)/3$ & $(2-x)/3$\\
  $q_{2,L}$     & $1/3$ & $1/3$  & $0$      &  $1/3$ & $1/3$\\
  $c_R$         & $1/3$ & $-1/3$ & $-y/3$   &  $y/3$ & $x/3$\\
  $s_R$         & $1/3$ & $-y/3$ & $1/3$    &  $(2-y)/3$ & $(2-x)/3$\\
  $q_{3,L}$     & $1/3$ & $1/3$  & $0$      &  $1/3$ & $1/3$\\
  $t_R$         & $1/3$ & $-1/3$ & $2\!-\!\frac{2}{3}(x\!\!+\!y)\!\pm\!\!\sqrt{3\!-\!x^2\!\!-\!y^2}$   &  $z/3$ & $x/3$\\
  $b_R$         & $1/3$ & $3+\frac{x+y}{3}$ & $1/3$    &  $(2-z)/3$ & $(2-x)/3$\\
  \hline
  $(\nu^e_L, e_L)$      & $-x$     & $x/3$  & $(x-1)/3$&  $-1$ & $-1-2y$\\
  $e_R$                 & $-x$     & $-1/3$ & $x/3$    &  $-(2+x)/3$ & $-(2\!+\!x)/3-2y$\\
  $(\nu^\mu_L,\mu_L)$   & $-y$     & $y/3$  & $(y-1)/3$&  $-1$ & $y-1$\\
  $\mu_R$               & $-y$     & $-1/3$ & $y/3$    &  $-(2+y)/3$ & $-(2\!+\!x)/3+y$\\
  $(\nu^\tau_L, \tau_L)$& $x+y-3$  & $3+\frac{x+y}{3}$  & $\frac{2}{3}-\frac{1}{3}(x+y)$ &  $-1$ & $y-1$\\
  $\tau_R$              & $x+y-3$  & $-1/3$ & $x\!+\!y\!-\!3\!\mp\!\frac{4}{3}\sqrt{3\!-\!x^2\!-\!y^2}$    &  $-(2+z)/3$ & $-(2+x)/3+y$\\
  \hline\hline
\end{tabular}
\end{table}
\section{Summary}\label{Sum}

In this paper, we have classified various new physics models involving
the $Z'$ boson in two different ways: one according to $Z'$
boson mixings with $Z$ and $\gamma$, and the other according to
$Z'$ boson charges with respect to quarks and leptons. In regard to the former,
we based  the general description for the
$Z'\!-\!Z\!-\!\gamma$ mixing derived from the EWCL on our previous
work\cite{Z'our}, characterizing these new physics models into five classes: 1.
Models with minimal $Z'\!-\!Z$ mass mixing; 2.Models with
minimal $Z'\!-\!Z$ kinetic mixing; 3.Models with general
 $Z'\!-\!Z$ mixing; 4.Models with $Z'\!-\!\gamma$ kinetic and
$Z'\!-\!Z$ mixing; and 5.Models with Stueckelberg-type mixing.
Although the general $Z'\!-\!Z\!-\!\gamma$ mixing is complicated and
there is no exact analytical expression for the mixing matrix $U$ and
masses $M_Z,M_{Z'}$, we
obtain explicit analytical expressions for each of our five simplifying classes. We find that the
most elementary mixing is the minimal $Z'\!-\!Z$ mass mixing, the
other four classes of mixings can be transformed into the minimal
$Z'\!-\!Z$ mass mixing through field transformations. In regard to the latter classification,
we exploit the anomaly cancellation conditions to constrain the $U(1)'$ charges. For generation-independent $U(1)'$ charges,
there are six charges
$y'_q$,$y'_u$,$y'_d$,$y'_l$,$y'_e$,$y'_\nu$ for which anomaly cancellation
requires that only two are independent parameters while the other
four can depend on these two parameters in two different
ways. While one appears already in the literature, the other is
new. For generation-dependent $U(1)'$ charges, we have listed
some  possible special solutions.

\section*{Acknowledgments}

This work was  supported by National  Science Foundation of China
(NSFC) under Grant No. 10875065.




\begin{thebibliography}{1}\label{biblio}

\bibitem{RizzoARXIV2006}
T.G. Rizzo, arXiv:hep-ph/0610104 (2006)

\bibitem{Hill}
C.T.Hill, Phys.Lett. B{\bf 345},483(1995).

\bibitem{Lane}
K.Lane and E.Eichten, Phys. Lett. B {\bf 352}, 382(1995).

\bibitem{Chiv}
F.Braam, M.Flossdorf, R.S.Chivukula, S.D.Chiara, E.H.Simmons, Phys.
Rev. D{\bf 77}, 055005(2008).

\bibitem{Z'SUSY}
P.Langacker, G.Paz,L-T Wang and I.Yavin, Phys. Rev. Lett. {\bf 100},
041802(2008); Phys. Rev. D{\bf 77}, 085033(2008).

\bibitem{CasselARXIV2009}
S. Cassel, D.M. Ghilencea, G.G. Ross, arXiv:0903.1119 (2009).

\bibitem{Langacker}
P.Langacker, e-Print: arXiv:0801.1345.

\bibitem{Z'our}
Y.Zhang, S-Z.Wang, Q.Wang, JHEP03,047(2008).

\bibitem{FranziniPRD1987}
P.J.Franzini and F.J.Gilman, Phys. Rev. D{\bf 35}, 855(1987).

\bibitem{RizzoPRD1991}
T.G.Rizzo, Phys. Rev. D{\bf 44}, 202(1991).

\bibitem{LangackerPRD1992}
P.Langacker, M.Luo, Phys. Rev. D{\bf 45}, 278(1992).

\bibitem{ChiappettaPRD1996}
P.Chiappetta, J.Layssac, F.M.Renard, C.Verzegnassi, Phys. Rev. D{\bf
54}, 789(1996).

\bibitem{FramptonPRD1996}
P.H.Frampton, M.B.Wise, B.D.Wright, Phys. Rev. D{\bf 54},
5820(1996).

\bibitem{ErlerPRL2000}
J.Erler, P.Langacker, Phys. Rev. Lett. {\bf 84}, 212(2000).

\bibitem{AnokaNPB2004}
O.C.Anoka, K.S.Babu, and I.Gogoladze, Nucl. Phys. B{\bf 687},
3(2004).

\bibitem{KozlovPRD2005}
G.A. Kozlov, Phys. Rev. D{\bf 72}, 075015(2005).

\bibitem{BassoARXIV2008}
L.Basso, A.Belyaev, S.Moretti, C.H.Shepherd-Themistocleous,
arXiv:0812.4313(2008).

\bibitem{ChanowitzARXIV2008}
M.S.Chanowitz, arXiv:0806.0890v2.

\bibitem{AppelquistPRD2003}
T.Appelquist, B.A.Dobrescu, and A.R. Hopper, Phys. Rev. D{\bf 68},
035012 (2003).

\bibitem{FerrogliaAP2007}
A.Ferroglia, A.Lorca, J.J.van der Bij, Annal. Phys. 16, 563(2007).

\bibitem{CarenaPRD2004}
M.Carena, A.Daleo, B.A.Dobrescu, and T.M.P.Tait, Phys. Rev. D{\bf
70}, 093009(2004).

\bibitem{RizzoARXIV1998}
T.G.Rizzo, Phys. Rev. D{\bf 59}, 015020(1998).

\bibitem{LangackerPRD2008}
P.Langacker, G.Paz, L.-T.Wang, I.Yavin, Phys. Rev.D{\bf 77},
085033(2008).

\bibitem{LangackerPRL2008}
P.Langacker, G.Paz, L.-T.Wang, I.Yavin, Phys. Rev. Lett.{\bf 100},
041802(2008).

\bibitem{Holdom1986}
B. Holdom, Phys. Lett. B{\bf 166}, 196(1986).

\bibitem{PDG2006}
W.-M. Yao et al, J. Phys. G{\bf 33}, 1(2006).

\bibitem{BabuPRD1996}
K.S. Babu, C. Kolda, and J.March-Russell, Phys. Rev. D{\bf 54},
4635(1996).

\bibitem{Hill1}
H-H.Zhang, S-Z.Jiang, J-Y.Lang, Q.Wang, Phys. Rev. D{\bf 77},
055003(2008)

\bibitem{Lane1}
J-Y.Lang, S-Z.Jiang, Q.Wang, Phys. Rev. D{\bf 79}£¬015002(2009).

\bibitem{Chiv1}
J-Y.Lang, S-Z.Jiang,  Q.Wang, Phys. Lett. B{\bf 673}, 63(2009).

\bibitem{HoldomPLB1991}
B. Holdom, Phys. Lett. B{\bf 259}, 329(1991).

\bibitem{KorsJHEP2005}
B.K\"{o}rs and P. Nath, JHEP 07, 069(2005).

\bibitem{FeldmanPRL2006}
D.Feldman, Z.Liu, and P.Nath, Phys. Rev. Lett.{\bf 97},
021801(2006).

\bibitem{FeldmanPRD2007}
D.Feldman, Z.Liu, P.Nath, Phys. Rev. D{\bf 75}, 115001(2007).

\bibitem{PDG2008}
C.Amsler et al., Phys. Lett. B{\bf 667}, 1(2008).

\bibitem{GUTs}
J.L. Hewett,T.G.Rizzo, Phys. Rept. {\bf 183}, 193(1989);  A.Leike,
Phys. Rept. {\bf 317}, 143(1999); P.Langacker, Phys. Rept. {\bf 72},
185(1981).

\bibitem{DavidsonPRD1979}
A.Davidson, M.Koca and K.C.Wail, Phys. Rev. D{\bf 20}, 1195(1979).

\end{thebibliography}
\end{document}